\documentclass[pre,twocolumn]{revtex4-1}
\usepackage[utf8]{inputenc}
\usepackage{graphicx}
\usepackage{hyperref}
\usepackage{xcolor}
\newcommand{\changed}[1]{ #1}

\begin{document}
\title{Numerical study of buoyancy induced arrest of viscous coarsening}
\author{Hervé Henry}
\email{herve.henry@cnrs.fr}
\affiliation{Laboratoire PMC, École Polytechnique, IPP, CNRS, 91128 Palaiseau, france}

\begin{abstract}
  The effect of buoyant forces on viscous coarsening is studied numerically. It is shown that at any time buoyant forces induce a vertical flow that scales like the Stokes velocity. This does no induce any  noticeable  change in the morphology of the coarsening microstructure under a  value of the characteristic length of the pattern. Above this threshold the pattern evolves toward a quasi two D pattern and coarsening stops. The characteristic length is shown to scale like $\sqrt{\gamma/(g \Delta\rho)}$ where $\gamma$ is he surface tension and  $\Delta\rho$ the mass dnsiy difference between the phases.
\end{abstract}
\maketitle
\section{Introduction}
In multiphase material  the specific spatial organization of the phases  affects dramatically material properties. This is illustrated for instance by the nacre where the arrangement  of soft and hard phases leads to a fracture toughness much higher than the fracture toughness of bulk materials\cite{Okumura2001,Hutchinson2001}. Another example are metamaterials where the spacial organization of the phases  leads to new physical properties\cite{Salman2021,Lemoult2013}. In both these examples the structure of the material that leads to peculiar properties has been \textit{designed}. However in most man made materials, it  arises through self organization during the manufacturing process\cite{MartinezFernandez2003}.  Among the mechanisms that lead to self organization, phase separation followed by coarsening is ubiquitous\cite{CahnHilliard}: an homogeneous mixture of two (or more) chemical species is stable at \textit{high} temperature due to entropic effects. When the temperature is decreased it is no longer stable and the mixture phase separates in regions of different compositions through diffusion. This phase separation takes place at small  lengthscales. It is followed by a coarsening regime\cite{LSW61,Cahn1965} during which the characteristic lengthscale of the pattern, $l$, increases through diffusion proportionally to $t^{1/3}$. If both phases are liquid this diffusive coarsening can be followed by a viscous coarsening\cite{Siggia1979} characterized by a linear increase of $l$ with time. This viscous coarsening process can only take place in a bicontinuous microstructure and is due to the breakup of capillary bridges of one phases that are part of a percolating network. After a breakup, the two protrusions created by this event retract and contribute to the reinforcement of other bridges that are on average thicker and longer. 

In the case of diffusive coarsening it has been shown that the details of the kinetics affect the microstructure dramatically\cite{Voorhees2020}. This is also the case of  hydrodynamic coarsening as it has been shown both experimentally\cite{Bouttes2014,Bouttes2015,Bouttes2016} and numerically\cite{Henry2018,Henry2019}. In all these cases, the microstructure is isotropic as can be expected since there is no symmetry breaking with respect to orientation. In actual systems this is often not the case since both phases do not share the same density (except in very specific cases) which leads to a symmetry breaking due to  buoyant forces.  The question that arises is when this symmetry breaking translates into a significant change in the microstructure and how it translates. It is expected that the buoyant forces  will lead to sedimentation  and the formation of a heavy-phase rich region at the bottom (light-phase rich region at the top). However, how the pattern  changes in the bulk between these boundary regions is unclear and the rate at which the sedimentation process will take place is not easy to estimate. Considering the existence of a threshold wavelength in the Rayleigh Taylor instability, it is  even possible that  below  a threshold lengthscale  there is no global flow.  In a different context, the sedimentation of  isolated inert monodisperse particles there is no  apparent modification of the microstructure\cite{Bergougnoux2009} in the absence of bottom and top wall interaction.  In the context of hydrodynamic coarsening of bicontinuous microstructure,  the superimposition of a plane Couette flow leads to strongly anisotropic patterns\cite{Beyssens,Berthier2001} but the flow is imposed externally. Hence, despite the fact that  have dramatic effects on material properties, the influences of buoyant forces or body forces such as magnetic ones\cite{LI2022,LI2020} on the microstructure during viscous coarsening  are not well understood yet.  
  
  Here, we present numerical results about  the pattern changes induced by buoyant forces on a coarsening bicontinuous structure. The results show that first there is a clear vertical macroscopic flow that does not lead to any visible microstructure changes. This stage is followed by a regime where the pattern becomes strongly anisotropic and where the coarsening velocity is dramatically reduced.  The remaining of the paper is organized as follows. First we present briefly the mathematical model that is used and the numerical methods together with the method of analysis that are used. Thereafter we present and  discuss the numerical results and conclude. 
  
\section{Model and numerical method}

The model used in this study is a phase field model\cite{kim_2021} where the interface between the phases is not tracked explicitly but is defined implicitly as an isosurface of an indicatrix function. This class of models  has become popular in the multiphase flow community and is now widely used\cite{Dadvant2021}. The model used here  has already been used in our study of the Rayleigh Taylor instability and is discussed in details in \cite{Zanella2020} where the convergence of the model toward a biphasic fluid  is shown and the scaling of the model parameters to achieve the fastest convergence is discussed in relation with previous theoretical studies\cite{Magaletti2013}.  So, we  limit ourselves to a rapid description  of the equations and of the effects of the model parameters. The equations are usually named the Cahn-Hilliard Navier-Stokes model and write~: 
\begin{eqnarray}
     \mathcal F&=& \int{ dV\,  G(c)+\frac{\epsilon}{2} (\nabla c)^2 }\label{eq_potentiel}\\
    \mu&=&\frac{\delta \mathcal{F}}{\delta c}=G'(c)-\varepsilon \Delta c\label{eq_mu}\\
	\partial_t c+ \mathbf{v.\nabla} c & =& \nabla (M \nabla \mu) \label{eq_advectiondiffusion}\\
		\partial_t \mathbf{v}+ \mathbf{v.\nabla} \mathbf{v} & =& -\frac{1}{\rho}\mathbf{\nabla} P +\nu \Delta\mathbf{ v} +\frac{\Delta \rho c}{\rho}\mathbf{g}-\frac{1}{\rho}c\nabla \mu \label{eq_NS}\\
    	\nabla v&=&0\label{eq_incompressible}
\end{eqnarray}
\changed{In eq.\ref{eq_potentiel} the expression for the total free energy $\mathcal{F}$ of the system is given as in Ref. \cite{CahnHilliard}. It is a function of the concentration $c$ of one chemical component. The  exact form of the function $G(c)$ and the value of $\epsilon$ affect the properties of the interface between two phases~: the surface tension $\gamma$ and the interface thickness $w_{int}$ (see table \ref{table_parameters}). The chemical potential $\mu$  that derives from this total free energy is   given in \ref{eq_mu}. And finally the advection diffusion equation is \ref{eq_advectiondiffusion} where $\mathbf{v}$ is the fluid velocity. The last two equations are the fluid flow equations. Eq.\ref{eq_incompressible} expresses the incompressibility condition while the eq.\ref{eq_NS} is the Navier-Stokes equation using the Boussinesq approximation. $\nu$ is the fluid velocity that is iindependant of the concentration $c$ The  source terms, $c\nabla\mu/\rho $ and $\Delta \rho c g$ correspond to the osmotic pressure ( once integrated through the  interface it  provides  the Laplace pressure jump) and to the buoyant force. Here,  both phases share the same viscosity.  The details of the model parameters are given in table \ref{table_parameters}.} This set of equation is solved numerically using a semi-implicit pseudospectral algorithm that is described in \cite{Henry2018} \changed{and briefly recalled in appendix}. \changed{The use of a spectral method  implies periodic boundary conditions in all directions and that the pattern characteristic lengthscale is limited by the box size. In the simulations presented here there are a few structure in each direction and the results should not be affected by finite size effects. This was confirmed by running simulations in twice larger and twice smaller boxes that lead to quantitatively similar results.} . The initial condition is, an already phase separated mixture with a volume fraction of the minority phase ranging from 0.35 to 0.5 as described in \cite{Henry2018}.     In the following we  discuss the case 0.35 because  in this case the topological changes are more visible. Before turning to the discussion of  numerical results, we find it necessary do discuss the different  parameters that are present in the model.
 \begin{table*}
 	\begin{tabular}{|c|c|l|}
 		\hline
 		Symbol        & value                          & description                                                      \\ \hline
 		$\rho$        & $ 1000\ \mathrm{kg/m}^3$        & inertial mass density                                            \\
 		$\rho(c)$     & $\rho+c\times \Delta \rho $    & buoyant mass density                                             \\
 		$\Delta \rho$ & 0.0625 to 64 $\mathrm{kg/m}^3$ &                                                                  \\
 		$M$           &                                & Mobility                                                         \\
 		$G(c)$        & $Ac^2(1-c)^2$                  & free energy functional                                           \\
 		$A$           & $6\gamma/w_{int}$              &                                                                  \\
 		$\epsilon$    & $3w_{int}\gamma$               & coeficient of the squared gradient term in eq.\ref{eq_potentiel} \\
 		$w_{int}$     & $\sqrt{2\epsilon/A}$           & equilibrium interface thickness                                  \\
 		$\gamma$      &0.005=$\sqrt{A\epsilon/18}$          & Surface tension                                                  \\
 		$\nu$         &                                & kinematic viscosity                                              \\
 		$v_s$         & $\gamma /(\rho \nu)$           & Siggia's coarsening velocity                                     \\ \hline
 	\end{tabular}
 	\caption{\label{table_parameters} List of parameters used in the simulations and other useful quantities}
 \end{table*}
 
 First, the parameters $A$ and $\epsilon$ are  present in the potential eq. \ref{eq_potentiel} through the prefactor of the gradient squared term and through parameters of $G(c)$ that is given  in table \ref{table_parameters} . They can be combined to give both a lengthscale and an energy scale that correspond to the interface thickness  $w_{int}\sqrt{2\epsilon/A}$ between the two equilibrium phases $c=0$ and $c=1$ and to  corresponding surface the surface tension $\gamma= \sqrt{A\epsilon/18}$. The effect of the mobility $M$ is more complex.  
  As discussed in \cite{Zanella2020}, it enters in the Péclet number that compares the velocity of the coarsening process induced by diffusion to the velocity of coarsening that is induced by interface motion and was first introduced in \cite{Siggia1979} $v_{S}\propto \gamma/(\rho\eta)$.  In the case where the diffusive transport is neglectable the mobility  must be chosen small enough to ensure that diffusion transport can be neglected at large scales. It must also be chosen large enough to keep the interface profile close to its equilibrium shape despite fluid flow induced deformation. We have used the parameter values that were found to be optimal in \cite{Zanella2020} (for each  given flow condition using as reference velocity $v_S$ and for characteristic length, the initial size of the pattern).  Since the Péclet number is varying during coarsening  we have  checked that changing the mobility by a factor of 2 and 1/2  did not lead to any measurable  change. This was expected since the relative error was found very close to the minimal error for a wide range of mobility values in \cite{Zanella2020}. 
  
  The influence of gravity terms can be measured through ,  another characteristic velocity: the Stokes sedimentation velocity that is a function of the characteristic lengthscale  of the microstructure and of the viscosity. It scales like $R^2 \Delta \rho/ (\rho \nu) $ and increases with the characteristic lengthscale. 
 
In our simulations surface tension is fixed, so that the Siggia's coarsening velocity is kept constant. The parameter that are  varied are the initial pattern characteristic lengthscale (together with the interface thickness to keep the Cahn number $w_{int}/l$ unchanged and the mobility, to keep the Péclet number unchanged).  The magnitude of the buoyancy force is changed  through the variation of $\Delta \rho$.  

 In order to quantify the pattern observed here a few quantities are  used and  must be defined. First in order to derive a characteristic lengthscale we use simply the ratio of the total volume over the surface area of the interfaces and in order to evaluate the later we use the ratio of the surface tension with the integral of the squared gradient of $c$ multiplied by $\epsilon$ which  is actually the surface energy of a plane interface  at equilibrium.
  \begin{equation}
     l=\frac{V}{ (2 \int \frac{\epsilon}{2} (\nabla c)^2)/\gamma}
    \end{equation} 
  This lengthscale is independent of the interface thickness. This contrasts with  the length defined  the ratio of the second and the third moment of the power spectrum which is  affected by the interface thickness. In addition in order to characterize the geometry of the pattern and its morphological changes we consider  the probability distributions of the principal curvatures of the interface rescaled by the characteristic length\cite{Kwon2007,Kwon2009,Kwon2010}. It is sensitive to pattern changes invisible to the naked eye and can be used to better understand the evolution of the microstructure.

 The flow  is solely characterized by the flow rate of one phase (and the other) long the three axis $x,\ y,\ z$. It should be equal to 0 if the flow is isotropic. Here since there is an obvious difference  between the z axis and the other 2 axis we choose to consider the vertical flow rate $Q_\parallel$~: 
 \begin{equation}
   Q_\parallel= \int_V \mathbf{v}.\mathbf{e_z}(1-c)\mbox{ or  } \int_V \mathbf{v}.\mathbf{e_z}(c)\label{eq_qpar}
 \end{equation}
 In the same spirit the horizontal flow rates are defined for $i\in\{x,\ y\}$ as:
\begin{equation}
   Q_\perp= \int_V \mathbf{v}.\mathbf{e_i}(1-c)\mbox{ or  } \int_V \mathbf{v}.\mathbf{e_i}(c) \label{eq_qperp}
 \end{equation}
The choice between the two expressions depends wether one is interested in the flow rate of the heavy or of the light phase.

As in \cite{Henry2019}, we also consider the conductance of the microstructure along a given orientation when one phase is conductive and the other not. Here  we consider both $G_\perp$ and $G_\parallel$ depending on whether the considered direction is perpendicular or parallel to the gravity field.  This quantity gives a measure of the  connectivity changes of the pattern without a strong  sensitivity on small structures that can be seen when considering topological invariants such as the genius number.   And furthermore we  only present the evolution of the conductance of the minority phase since it is more sensitive to changes of the microstructure.  

\begin{figure*}[t]
	\begin{tabular}{ccc}
		\includegraphics[width=0.33\textwidth]{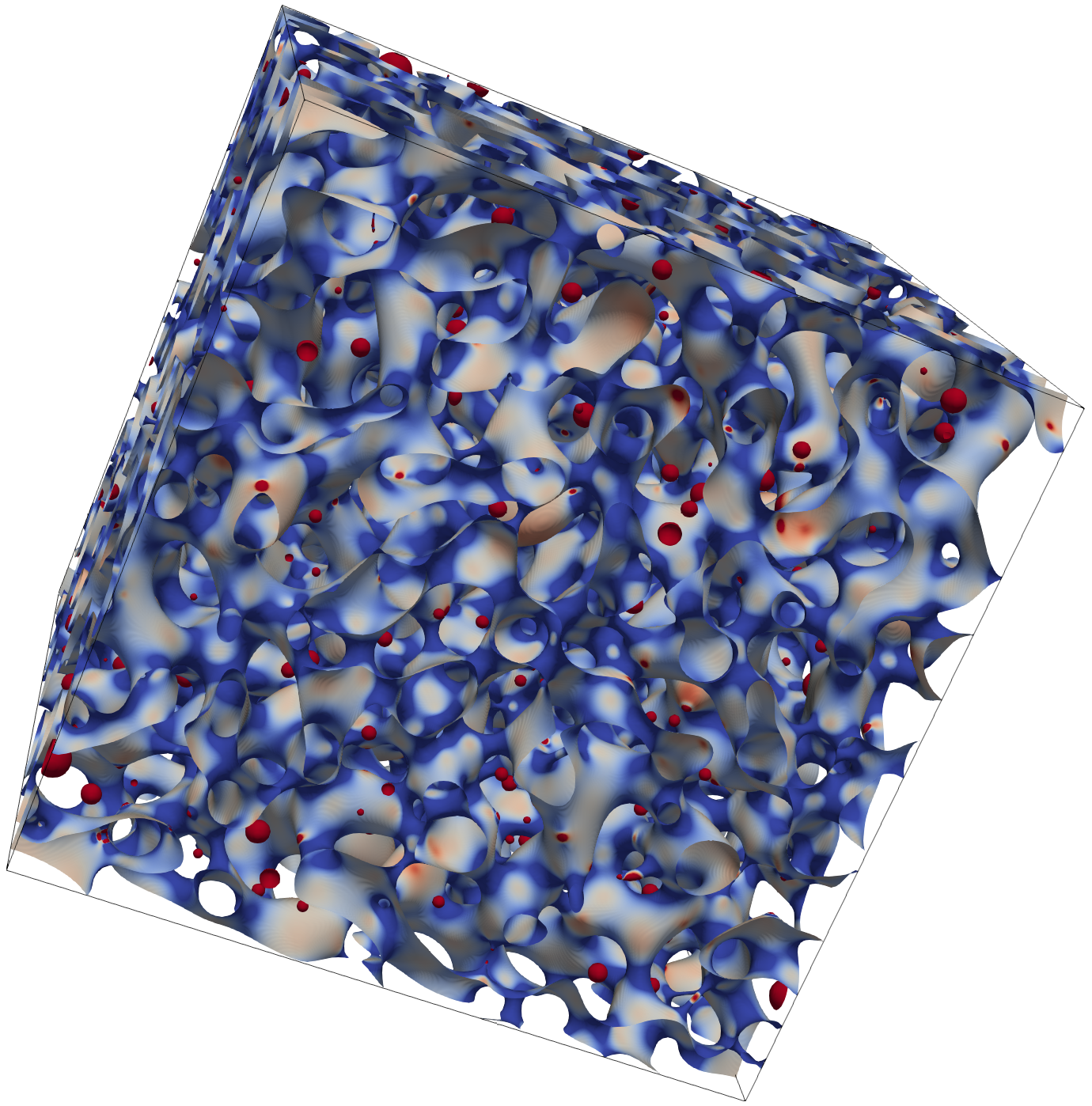}
		&
		\includegraphics[width=0.33\textwidth]{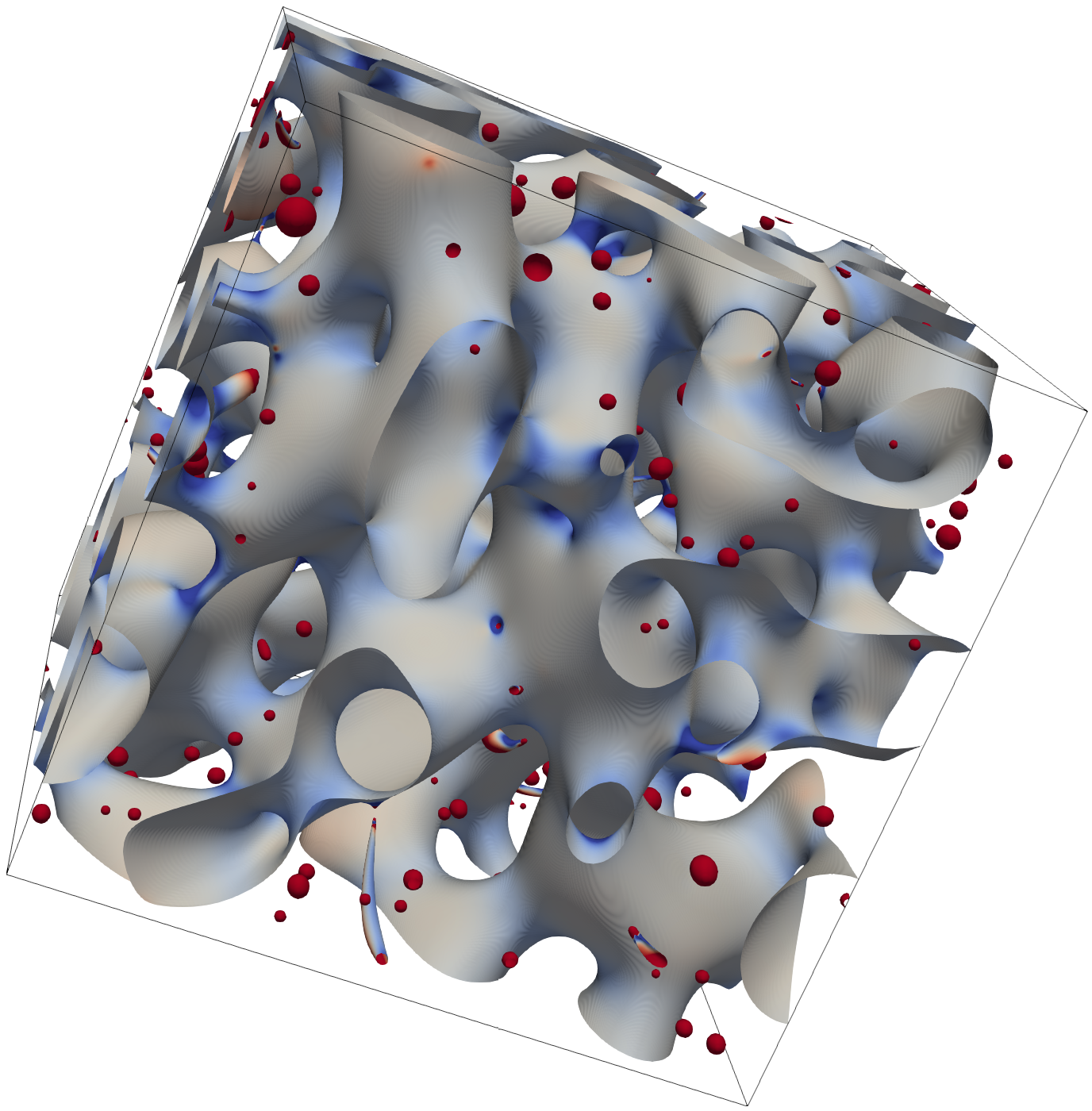}
		&
		\includegraphics[width=0.33\textwidth]{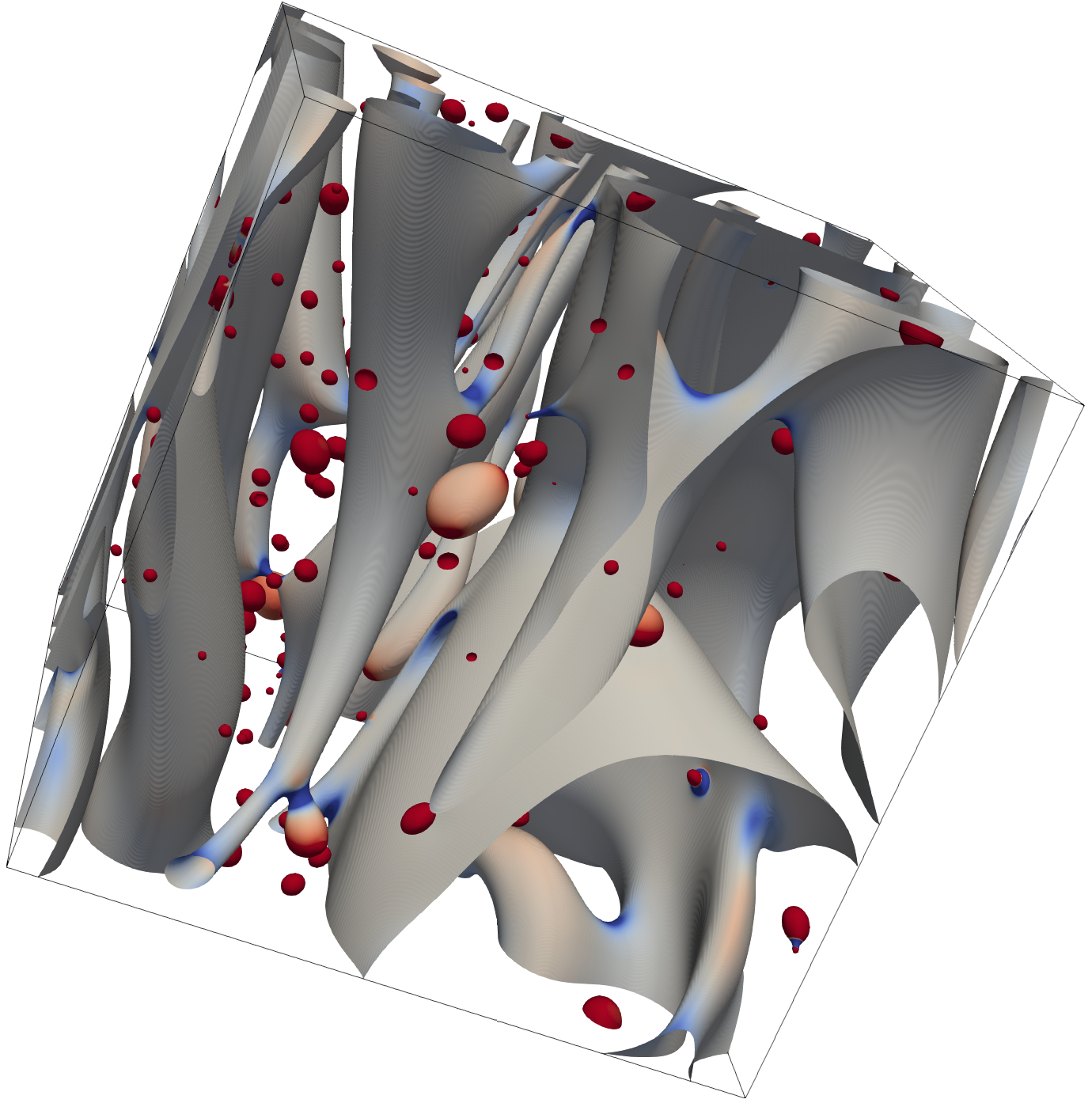}
	\end{tabular}
  \caption{\label{fig_snapshots}pattern during the gravity induced  slow down/arrest of the coarsening process. Here the volume fraction of the minority phase is 0.35. The interface between the two phases, \changed{that is defined as the isosurface $c=0.5$}  is colored with the gaussian curvature in order to make more apparent the numerous bubble that are present. In the absence of a buoyant driving force, there are not such bubbles.}
\end{figure*}

\section{Results}
\begin{figure}
	\includegraphics[width=.4\textwidth]{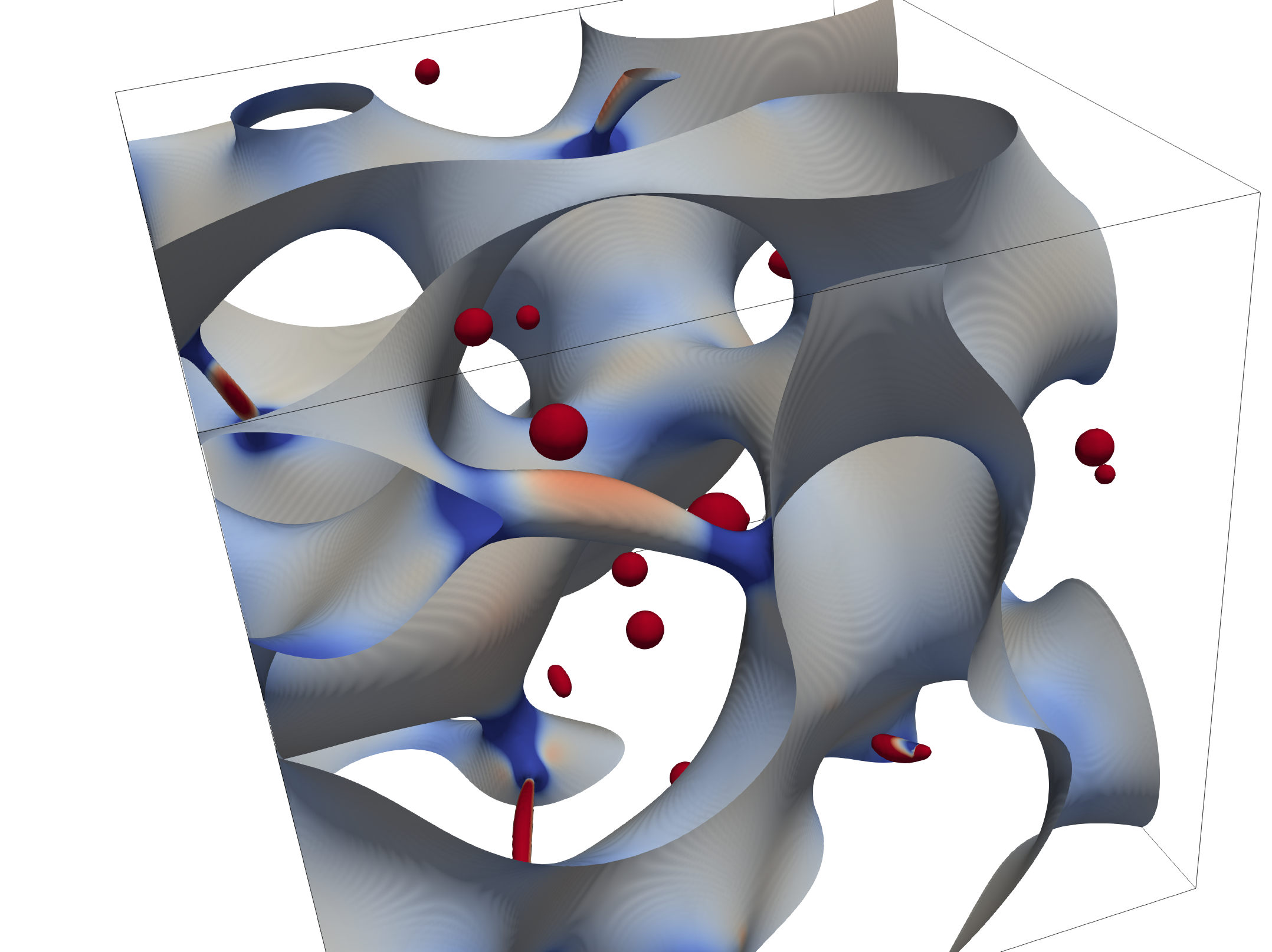}
	\caption{\label{fig_detail}Detail of the structure in fig.\ref{fig_snapshots} where one can see a capillary bridge close to breakup. The presence of two necks is due to the vertical flow that pulls on the filament and favors breakup at the ends, leading to the formation of bubbles. This contrasts with the situation in the absence of buoyant flow where  the bridges tend to breakup at the middle.}
\end{figure}
We now turn to a description of our results. To begin we give a qualitative description of the microstructure changes. Thereafter the effects are described more quantitatively.

First, our simulations show that the buoyant term can induce a pattern change that is not related to bottom (or top) hard wall effects. Indeed it was found while for low buoyant forcing  it is nearly impossible to measure any effect, if $\Delta \rho$ is large enough  the microstructure is affected.  
In this case, the typical evolution of the pattern  can be described as follows. After a self similar regime where the effects buoyant forces are almost neglectable, there is an increase of the anisotropy of the pattern that keeps its bicontinuous nature. Interfaces along the gravity axis are becoming more important while structures transverse to the gravity axis tend to vanish. This results in the formation of a quasi-2D pattern that is almost  no longer bicontinuous. It consists of layers of on phase or the other that are mainly oriented along the z axis and that are flowing vertically as can be seen in fig. \ref{fig_snapshots}. It  induces an arrest  of viscous the coarsening process \changed{at a critical lengthscale $l_{max}$  that is the characteristic  lengthscale of the 2D pattern}. The later evolution of the microsctructure is out of the scope of this work.  This behavior is illustrated in figure  \ref{fig_snapshots} where the initial microstructure is represented, followed by two patterns obtained at later times. While the only difference between the first two pattern seems to be the characteristic length, there is a dramatic change when considering the last picture where the interfaces are mostly vertical. During this evolution one can also notice the presence of multiple bubbles (whose presence is made more obvious through a coloring of the interface with the gaussian curvature). Their origin  is illustrated in figure \ref{fig_detail}. Indeed, in this figure, a small region of the simulation domain is represented and in this region a capillary bridge just before breakup is represented. This bridge is roughly orthogonal to the vertical axis and one can see that it presents two necks that correspond to the future breaking points. This will eventually lead to the breakup of both ends of the capillary bridge and the formation of a bubble. The formation of two  necks is due to the force exerted by the vertical flow of the on the capillary bridge that bends the filament. This  behavior is not observed in zero-gravity coarsening where the breakup takes place in the middle of the filament. It should be noted that during early stages of the coarsening the formation of small bubbles is a visible  specificity of  coarsening with  gravity. Less visible when looking at the microstructure is the vertical flow  that is present and can be measured through vertical flow of phase defined by eqs \ref{eq_qpar}, \ref{eq_qperp}. In the case considered here, due to the buoyant forces we expect the flow rate along the vertical axis, $Q_\parallel$ to differ from 0 while the flow rate along the horizontal directions $Q_\perp$ should remain 0. 
 \begin{figure}
	\includegraphics[width=.45\textwidth]{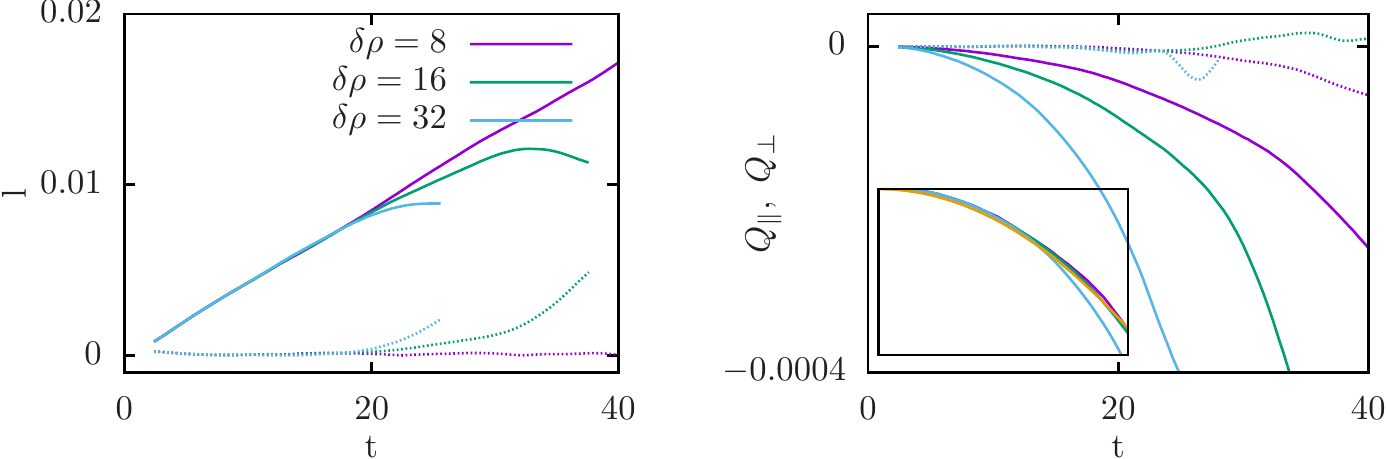}

	\caption{Evolution of the characteristic length (left solid ) and of the vertical flow rate of the heavy phase  (right solid). The dashed lines correspond to the difference with the linear fit that corresponds to the evolution of the characteristic length with the same parameters in the absence of buoyancy(right) and to  the horizontal flow rate in one horizontal direction (right). The averrage flow rate for the 20 last time units of the $\delta \rho=8$ case correspond to a motion of about 20\% of the computation domain (that is a few characteristic lengths of the pattern). In the inset, the flow rates rescaled by $\delta \rho$. \label{fig_length_flow} }
\end{figure}

In the following of this section the effects of the parameters is discussed. To this purpose we first consider a set of numerical simulations that share the same initial conditions and parameters with the exception of $\Delta \rho$ that is varied from 8 to 32.   In figure \ref{fig_length_flow}, one can see both the evolution of the characteristic lengthscales and of the vertical and horizontal flow rates  as a function of time. In all cases there is a well defined regime during which the characteristic lengthscale grows linearly with time (during this regime it is multiplied by a factor of at least 10). For  $\Delta \rho=8$ the linear regime spans the time range considered. This contrasts with the cases    $\Delta \rho=16$ and $\Delta \rho=32$ for which  the characteristic lengthscale stops growing after a finite time (or at a given value). Thereafter it  decreases. This  allows to define the coarsening arrest length as the maximal value of the characteristic lengthscale $l_{max}$. 
    The fact that the formation of a quasi 2D pattern corresponds to an arrest of the coarsening process is expected. Indeed, the viscous coarsening regime described by Siggia can only take place in a bicontinuous structure. Such pattern can be easily found in 3D systems but  they cannot exist in 2D. Hence once the pattern has reached a quasi bidimensional state, it cannot coarsen through  hydrodynamic coarsening mechanism. After the coarsening has stopped, the characteristic lengthscale of the pattern decreases. This corresponds to an increase of the interface surface that can be attributed to the formation of bubbles discussed above and hydrodynamic instabilities that are not discussed here. 
    
\begin{figure}
	\begin{tabular}{cc}
    \rotatebox{90}{$\rho=1008$}	&\raisebox{-0.5\height}{\includegraphics[width=0.4\textwidth]{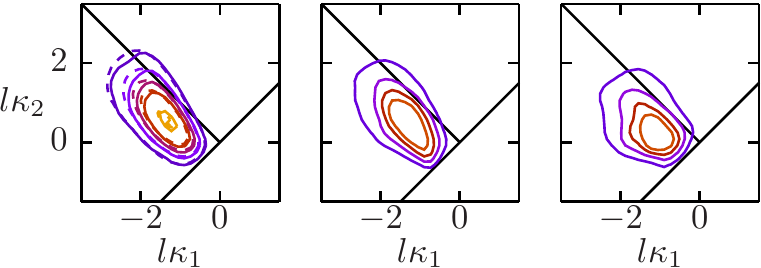}}\\
    \rotatebox{90}{$\rho=1016$}	&\raisebox{-0.5\height}{\includegraphics[width=0.4\textwidth]{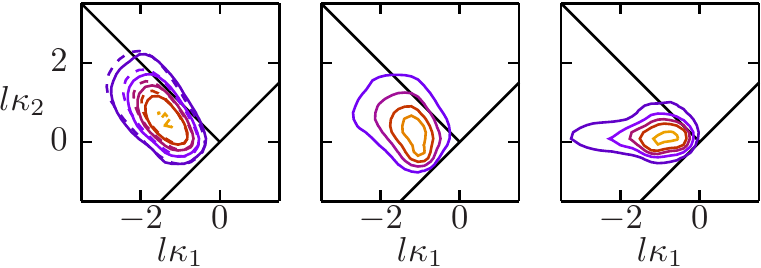}}\\
      \rotatebox{90}{$\rho=1032$}	&\raisebox{-0.5\height}{\includegraphics[width=0.4\textwidth]{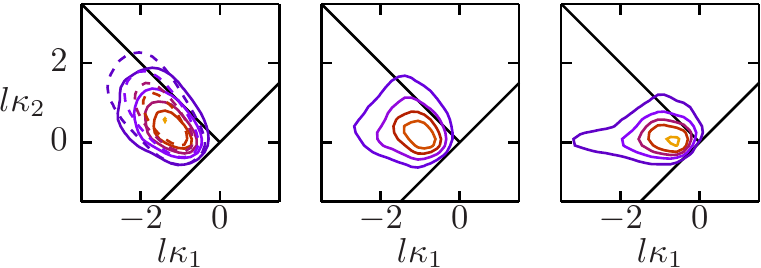}}
	\end{tabular}
  \caption{\label{fig_PDFS}Plot of the probability distributions of the principal curvature at different times for three different values of the mass density difference. On the leftmost plots, two times have been superimposed (dashed and solid lines) and the characteristic lengthscale has been multiplied by $\approx 2$ between these two instants.}
\end{figure}

    While below the maximal length (until $\approx l_{max}/2$) the buoyant forces have little visible effect on the evolution of the pattern, the flow is affected by the buoyant forces. Indeed, as illustrated in fig.\ref{fig_length_flow}  the vertical flow rate grows like $t^2$ (i.e. like $l^2$) which is consistent with the scaling of the Stokes sedimentation velocity. Hence there is a regime for which the vertical flow rate is present while the pattern is apparently unchanged.  This is confirmed by the study of the principal curvature PDFs. Indeed, one can see in fig. \ref{fig_PDFS} (a1,b1)  that, at the beginning of the coarsening process (the lengthscale has been multiplied by $\approx 2$   ), the self similar regime is still present: the contour lines taken at two different times superimpose very well after rescaling. However, at later times there is a visible change of the PDF indicating that the pattern is changing and the change is not due to the presence of bubbles that are out of the range of the plots  (high curvatures). There is a slight shift of the pattern toward the $l\kappa_2=0$ axis that corresponds to surfaces that contain straight lines such as planes, cylinders. Here  since the outcome of the pattern evolution is a similar to vertically aligned fluid sheets  we attribute this to the fact that the interfaces are remodelled by the flow and tend to align with it.

Finally, we discuss the evolution of conductivities with time. Indeed in both cases for which there is the formation of a quasi 2D pattern (only clearly visible at the very end of the evolution), one can see a slow and continuous decrease of $G_\perp$ with time that is consistent with the alignment of the pattern with the flow: the vertically aligned non conductive majority phase acts as a barrier to the flux (of electrons, of diffusing particles).   The fact that  there is a significant increase of the conductivity along the vertical axis is also consistent with this phenomenon. However  the complexity of fluxes in the irregular geometry \cite{havlin2002,Bouchaud1990} including   phase separated mixtures \cite{Barman2019,Barman2021} does not allow to be more specific. 
\begin{figure}
         	\includegraphics[width=.45\textwidth]{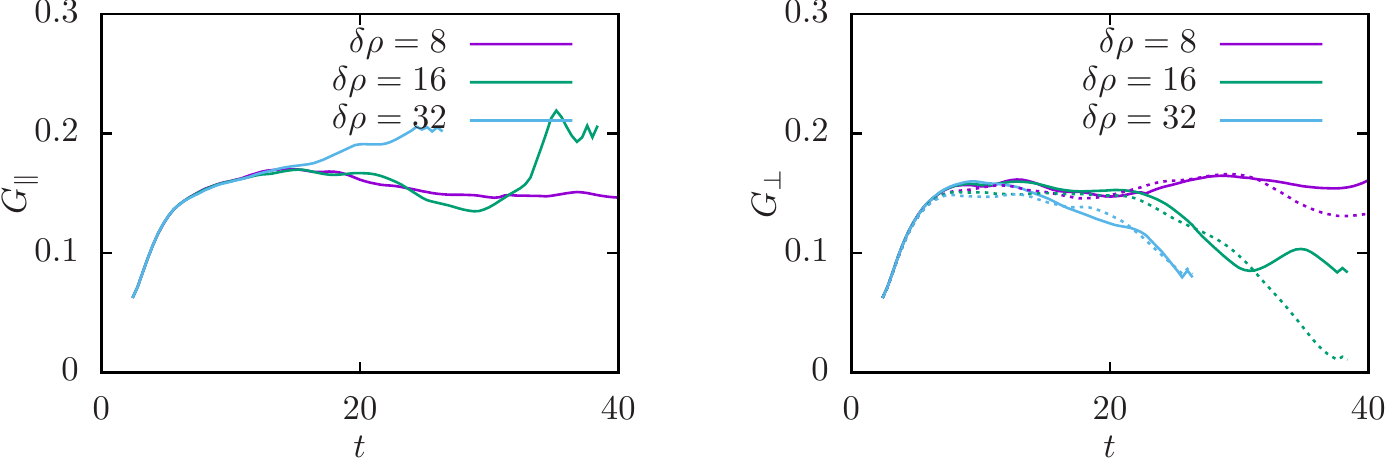}

	\caption{Evolution of the conductivities of the microstructure (assuming the minority phase is conductive and the majority phase is isolant) with time. On the left the conductivities along the z axis are plotted while on the righ the conductivities along the $x$ (solid) and $y$ (dashed) axis are plotted.}
\end{figure}

\begin{figure}
      \includegraphics[width=0.5\textwidth]{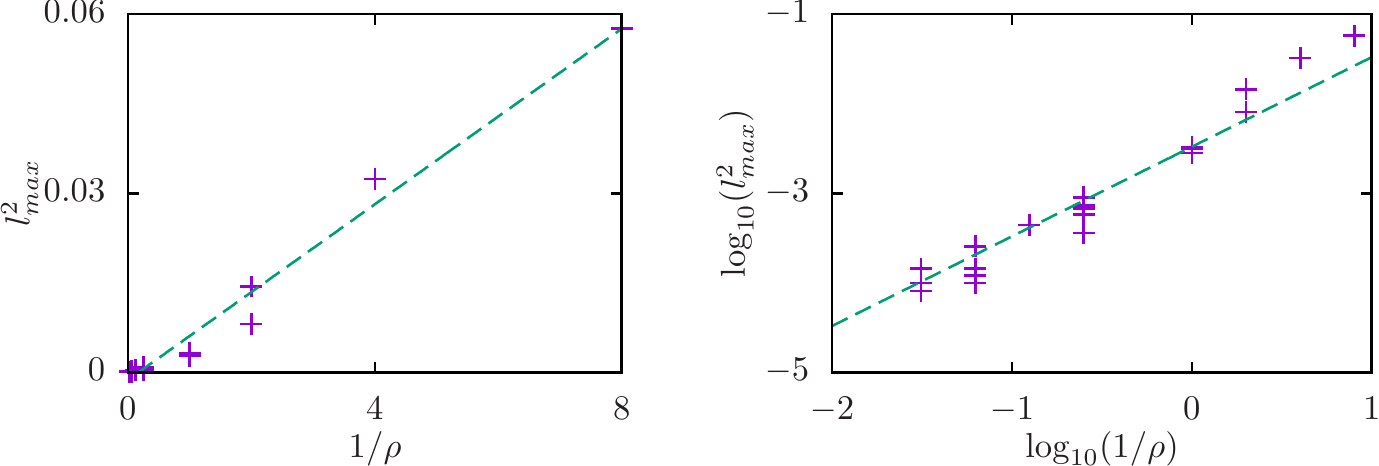}
      \caption{\label{fig_scaling} Plot of the  maximal lengthscale as a function of the density difference. In (a) a linear scale is used while in (b) a logarithmic scale is used. The points correspond to different initial conditions, with different characteristic lengthscale. All initial conditions used here are isotropic.}
    \end{figure}

From this we have a clearer view of the behaviour of an infinite coarsening medium  in the presence of buoyant forces. However, it is still unclear how the threshold length above which the quasi 2D pattern appears is scaling. From a comparison between the two characteristic velocities present in the system: the Siggia's coarsening velocity and  the Stokes sedimentation velocity one expects the length to scale as $\sqrt{\gamma/\delta \rho} $ and to be independent of the viscosity.   We have chosen to consider that the  lengthscale at which the transition takes place is the maximal lengthscale  that is reached during the coarsening. In figure \ref{fig_scaling} this length is ploted as a function of $\Delta \rho$ together with a fit using $A\sqrt{\delta \rho}$ and using a logarithmic scale together with linear fits with slope $0.5$ (the best fit with varying slope is  obtained for a slope of 0.53). One can see that the agreement is very good while initial conditions were varied (different characteristic lengthscales) and viscosity was also varied. This indicates that the expected scaling is verified at leading order: for a given parameter set the variations of the maximal length that was reached were at most of a factor 2 while the initial condition's characteristic lengthscale was varied by a factor up to 4. This was typically the case for $\Delta \rho =4$, parameter for which the changes in $l_{max}$ are at most of $20\%$.\changed{ From this it appears that the pattern keeps a three dimensional structure as long as the Stokes semdimentation velocity is significantly lower than the Siggia's coarsening velocity, which allows us to that below a critical lengthscale that scales like 
\begin{equation}
  l_{max}\propto \sqrt{\frac{\gamma}{\delta \rho}} 
\end{equation}
the pattern remains isotropic while above this critical lengthscale  it evolves toward a quasi two dimensional pattern and the coarsening process is almost stopped.  
}

\section{conclusion}
\changed{ Here the effect of buoyant forces on viscous coarsening has been studied in the Stokes regime and when the transport through diffusion can be neglected. It has been shown that there exists a characteristic lengthscale range for which  the microstructure is apparently unaffected despite the fact   there is a global vertical flow that has an average velocity of the order of the Stokes velocity. Above this characteristic length the microstructure dramatically changes: from an interconnected networks in 3D to  a quasi 2D pattern. This transition to a quasi 2D pattern induces an arrest of the hydrodynamic coarsening since the microstructure is no longer bicontinuous. 
Hence our results  bring light on  two aspects of the effects of  buoyant forces  on coarsening systems. First  an estimate of the flow rate as a function of both the density difference and the characteristic lengthscale of the pattern is given  when the microstructure is not visibly affected. Second  an estimate of the threshold lengthscale above which the microstructure is affected and becomes quasi 2D is given. This work can be of interest in the context of phase separation in industrial processes since it allows to actually estimate  the sedimentation rate and hence how the material will organize along the vertical axis. It also allows to predict when the bulk phase will loose isotropy. This could be used for instance to produce oriented microstructure that have different properties, such as conductance,  along different directions.   It must also be  noted that the computations presented here were performed with buoyant forces, but this can be easily extended to materials where the phases respond differently to external fields such as electromagnetic fields\cite{LI2020,Zanella2019}. }
\section{Aknowledgments}
This work was granted access to the HPC resources of IDRIS under the allocation AD012B07727R1 made by GENCI.\newline

\appendix

\changed{

\section{Numerical scheme}

Here we discuss briefly the numerical scheme used in our simulations. It is based on  a
Pseudo-spectral methods similar to the one that have  been used to compute turbulent flows \cite{Orszag_1969_PHYSFLUIDS,Orszag_1972_PRL}, and multi-phase flows \cite{Liu_2003_PHYSD} due to their high spatial accuracy(i.e. both amplitude and phase errors decaying exponentially with the resolution). The low dispersion characteristics of the  method ensures, that the subtle balance between inertial and viscous forces will not be modified by the numerical dissipation.
A further advantage of the method, that a numerically more stable solution is possible using the operator splitting technique at no extra cost: using backward Euler time integration for the viscous term, while forward Euler for the remaining terms. Besides is is straightforward to "force" incompressibility using the Helmholtz theorem:
projection to divergence-free velosity field does not require to explicitly compute pressure, that is used to force incompressibility.  
 
The evolution equation for the concentration is also solved using a similar approach: the linear high order term is solved using an implicit scheme in Fourrier space while non linear terms are computed in real space.

}\end{document}